\documentclass[12pt]{article}
\usepackage{times}

\usepackage{graphicx}

\newenvironment{sciabstract}
{\begin{quote} \bf}
{\end{quote}}

\newcounter{lastnote}

\title{Ultrafast switching to a stable hidden topologically protected quantum
state in an electronic crystal.}

\author
{
L. Stojchevska$^{1,2}$, I. Vaskivskyi$^{1}$, T. Mertelj$^{1}$,
P. Kusar$^{1}$, \\
D. Svetin$^{1}$, S. Brazovskii$^{4,5}$ and D. Mihailovic$^{1,2,3,*}$\\
\\
\normalsize{$^{1}$Dept. of Complex Matter, Jozef Stefan Institute, Jamova 39,
Ljubljana,
SI-1000, Ljubljana, Slovenia,}\\
\\
\normalsize{$^{2}$Jozef Stefan International Postgraduate School, Jamova 39,
Ljubljana SI-1000, Slovenia}\\
\\
\normalsize{$^{3}$CENN Nanocenter, Jamova 39, Ljubljana, Slovenia}\\
\\
\normalsize{$^{4}$LPTMS-CNRS, UMR8626, Univ. Paris-Sud, Bat. 100, Orsay, F-91405
France}\\
\\
\normalsize{$^{5}$International Institute of Physics, 59078-400 Natal, Rio Grande do Norte, Brazil
}\\
\\
\normalsize{$^\ast$To whom correspondence should be addressed; E-mail:  dragan.mihailovic@ijs.si.}
}

\date{07/04/2014}

\begin{document}


\baselineskip24pt

\maketitle

\begin{sciabstract}
Hidden states of matter with novel and unusual properties
may be created if a system out of equilibrium can be induced to follow
a trajectory to a state which is inaccessible or does not even exist
under normal equilibrium conditions. Here we report on the discovery
of a hidden (\emph{H}) topologically protected electronic state in a layered
dichalcogenide \emph{1T}-TaS$_{2}$ crystal reached as a result of a quench caused by a single 35 fs laser pulse.
The properties of the \emph{H} state are markedly different from any other state of the system: it exhibits a large drop of electrical resistance,
strongly modified single particle and collective mode spectra and
a marked change of optical reflectivity. Particularly important and unusual, the \emph{H} state is stable for an arbitrarily long time until a laser pulse, electrical current or thermal erase procedure is applied, causing it to revert to the thermodynamic ground state.
Major observed events can be reproduced by a kinetic model describing the conversion of photo excited electrons and holes into an electronically ordered crystal, thus taking into account a dynamic conversion of particles among reservoirs of liquids of electrons and holes, and of an electronically ordered crystal.
converting a  Mott insulator to a conducting \emph{H} state.
Its long-time stability follows from the topological protection of the number of periods in the electronic crystal.
\end{sciabstract}

Extreme conditions necessary for reaching hidden many body states
can be created in the laboratory using laser photoexcitation in condensed
matter systems. Typically,  the ground state ordering can be
temporarily destroyed, but on cooling the system reverts back in a few
picoseconds, exceptionally passing though a transient metastable state.
So far, a few such metastable states have been shown to
persist on timescales between $10^{-9}$ - 10$^{-3}$s ({\it 1,2,3,4,5,6,7})
before recovering to the ground state by a combination of thermal,
electronic and lattice relaxation processes ({\it 2}).
Stability of photoinduced states has been demonstrated in a manganite ({\it 6}) and in chalcogenide glasses ({\it 8}), where switching occurs
between neighbouring thermodynamic states, but no hidden states are
involved in these cases.

In this paper, we report for the first time on bistable switching
to a hidden ($H$), spontaneously ordered macroscopic quantum state whose
properties are distinct from any other state in the equilibrium phase diagram.
 The hidden state transition (HST) occurs in a layered quasi-2D chalcogenide
\emph{1T}-TaS$_{2}$ crystal, a system which exhibits multiple competing
ground states already under equilibrium conditions. Some of the states
are shown schematically in Fig. 1A.
Near $T_{c0}=$ 550 K \emph{1T}-TaS$_{2}$ forms an incommensurate
charge-density-wave (\emph{IC}) with an associated lattice distortion. On cooling these modulations sharpen to form star-shaped polaron clusters (Fig.1 A). Their ordering is thought to be responsible for a variety of phases, causing a transition to a nearly commensurate (\emph{NC}) state for $T<T_{c1}=350$ K, and a hysteretic first-order transition to a gapped commensurate (\emph{C}) phase at $T_{c2}\simeq183$ K.
On heating, the system develops a triclinic stripe-like ordered state
around 223 K which reverts to the \emph{NC} state at $T_{c2}\simeq 283$
K ({\it 9}). Further nearby equilibrium states are revealed upon the application of external pressure ({\it 10}) or doping ({\it 11}), both of which make \emph{1T}-TaS$_{2}$ superconducting.

To induce the HST, we use a single sub-35 fs Write
(W) pulse from an amplified Ti-Sapphire laser at 800 nm with energy
$U_{W}\simeq1$ mJ/cm$^{2}$.
After inducing a HST at 1.5 K, the 4-probe resistance $r(T)$ drops
approximately 3 orders of magnitude and \emph{remains in this state
indefinitely} at this temperature (Fig. 1B).
Upon heating, $r(T)$ is approximately constant up to 60 K, whereupon
it increases and merges to the virgin $r(T)$ curve corresponding to the C state above $T_{H}\sim100$
K. The \emph{I-V} characteristic remains linear throughout. Empirically
we found that the H state can be \emph{completely} erased (E) by a
pulse train of $10^{4}$ 50 ps pulses, each with $U_{E}\simeq1$ mJ/cm$^{2}$.
Alternatively, Joule heating can be used for erasure by passing a
$\sim0.1$ mA current through the device. In all three cases the system
reverts to the \emph{C} state. Stable switching can be achieved also
at intermediate temperatures up to $T\sim70$ K. The effect is entirely
reversible from cycle to cycle, sample to sample, irrespective of
the sample growth batch, and there appears to be no limit on the number
of W/E cycles that can be performed. (Experimental details on thermal
protocols, including ageing effects ({\it 12}),
and a description of the laser lithography used to manufacture the
contacts are given in the supplementary material (SM)).

To gain insight into the microscopic nature of the hidden state, we investigate the single-particle and collective excitations, using Pump-probe (\emph{P-p}) spectroscopy with the \emph{P} and \emph{p} pulse energies kept low ($<10 \mu$J/cm$^{2}$ and $<$1 $\mu$J/cm$^{2}$ respectively) to ensure minimal disturbance
of either state. The sample reflectivity $R(t)$ is simultaneously recorded by the probe (\emph{p}) beam. In Fig. 2 A to D we first present the transient reflectivity $\Delta R/R$ of
\emph{1T}-TaS$_{2}$. In the virgin \emph{C} state we observe oscillations due to the coherent excitation
of the amplitude mode (AM) and phonons which are superimposed on a background from exponentially decaying single particle (SP) excitations across the gap ({\it 13}). The spectrum
$S(\omega)$ obtained by Fourier transformation shows a strong amplitude mode at 2.46 THz, and weaker
phonon modes at 2.1, 2.18, 3.2 and 3.85 THz (see Fig.2B).
The HST modifies the $\Delta R/R$ (Fig. 2B) and the SP signal is reduced significantly.
In the spectrum after the HST shown in Fig.2B the AM peak at 2.46 THz
\emph{disappears}, in favour of a new mode at 2.39 THz; intensities
of modes at 3.10 THz and 3.85 THz are reduced, and some additional
spectral intensity appears between 2 and 2.5 THz.
On heating, the spectrum of the H state remains unchanged until $\sim70$K. Above 70 K it gradually reverts back to the \emph{C} state. Concurrent with the switching of the AM and phonons, we observe a switching of
reflectivity $R$ at 800 nm, as shown in Fig. 2 D.
All the observations display typical threshold behaviour as a function of $U_{W}$: below threshold fluence the resistivity, AM frequency
and reflectivity revert to the C state values after the W pulse. Close to
threshold fluence, the AM shows bimodal behaviour, which we interpret
as incomplete switching.
Importantly, no intermediate shift of the AM is observed, indicating distinct two-phase behaviour.
The \emph{H} state spectrum is quite different from the \emph{NC }state spectrum (Fig. 2 A and B) or the $T$ state spectrum (see SB), indicating that it is not related to the equilibrium states.

We emphasize some remarkable features of the HST:
\textbf{1.} After photo-excitation the \emph{H} state spontaneously orders below $T_{H}$, as indicated by the narrowness of the AM peak and the fact that \emph{no partial frequency shift} is ever observed even when incomplete switching is caused by near-threshold excitation (see SM).
\textbf{2.} The switching occurs \emph{only with short pulses}, and the threshold increases with increasing $\tau_{W}$ as shown in Fig. 2 C and - remarkably - can no longer be achieved with $\tau>4$ ps at any $U_{W}$.
\textbf{3.} The \emph{H} state is completely stable until erased, or heated above $\sim70$ K.
Note that $T_{H}$ has no special significance under equilibrium conditions
and is relevant only for describing the transition from the  \emph{H} state
to the \emph{C} state.

To understand these unusual phenomena, we first introduce a  scenario for switching based on the current understanding of the electronic ordering in $1T$-TaS$_{2}$ ({\it 9,14,15,16}), and then describe the results of a minimal phenomenological model which offers insight into how transient photodoping can lead to a stable ordered state.

The relevant electronic states  of \emph{1T}-TaS$_{2}$ in the C state which are within reach of our 1.5 eV laser photons are shown in Fig. 3 C. They are formed predominantly from a single Ta \emph{d} band which is split into subbands by the formation of a CDW depicted in Fig 1 A. Six of these subbands are filled with 12 electrons per new large unit cell forming a manifold of occupied states up to 0.4 eV below $E_{F}$ (shown in blue in Fig. 3 C). The 13$^{th}$ left-over electron is localised on the central Ta causing inward radial displacements of 12 neighbours in the shape of a star of David ({\it 14,15}), thus forming a self-localized polaron.
The 13$^{th}$ electron gives rise to a half-filled narrow metallic band straddling the Fermi level; but this band is further split by the Coulomb interaction into two Hubbard bands  (plotted green in Fig. 3 C) ({\it 15}), whereby  the upper one merges with the manifold of unoccupied subbands above $E_{F}$, while the lower one is $\sim0.2$ eV below $E_{F}$ - still well above the top of the valence band at $- 0.4$ eV, which makes the \emph{C} state  a Mott insulator in the form of a polaronic crystal ({\it 10,14,15}).
In the photoexcited case however, a charge imbalance can cause voids in the polaron lattice which - upon relaxation - leads to the formation of a spontaneously ordered \emph{H} state. The possibility of both positive and negative discommensurations has been indicated by STM experiments ({\it 9}), indicating that the system can accommodate both extra holes or extra electrons.

Recent experiments show that the Mott gap rapidly melts within $\sim50$ fs after photo-excitation ({\it 17}), but after $0.5\sim1$ ps, the AM oscillations are visible again, indicating the recovery of the \emph{C} state within a few AM cycles. Since the \emph{e-h} energy relaxation occurs on a similar timescale ($\tau_{E}\sim1$ ps) ({\it 13,17}), the condensation into the crystalline state competes with $e-h$ recombination.

Photoexcitation initially creates equal numbers of electrons ($e$) and holes ($h$) by an inter-band transition, followed by rapid intraband thermalisation via scattering amongst themselves and with the lattice, as well as transitions between different bands, reaching states near the Fermi level and melting the \emph{C} order on a timescale on the order of 50 fs ({\it 17,18,19,20}). The maximum effective electronic temperature reached in the process is $T_{e}\sim1000$ K, while the lattice reaches  $\sim150$ K within $3-5$ ps, whereupon the two are in quasi-equilibrium. (See SM for temperature measurements and model estimates). However, the large asymmetry of the band structure in this compound ({\it 14}), can also lead to a photodoping effect: the $e$ and $h$ scatter and lose energy at different rates, leading, on the sub-5-picosecond timescale, to a transient imbalance of their respective populations $n_e$ and $n_h$.

Let's now examine the photodoping effect in more detail. Doping away from half-filling of a conventional Mott-Hubbard state on a rigid lattice leads to a conducting state. But here photo-doping disturbs the polaronic deformations. Let's consider the effect of a photo-doped hole, which annihilates with the localised 13$^{th}$ electron. The annihilation removes the charge at the centre of the polaron, rapidly dissolving the polaronic distortion and leaving a void in its place. In the standard polaron picture ({\it 21}), its dissolution releases a band state from which the polaron was originally formed, which makes the system conducting.
Since some of the 13$^{th}$ electrons have been  annihilated by holes, not all ions in these regions are charge compensated, and they have an excess charge. Yet these regions cannot conduct because the remaining 12 electrons are in filled states within the gap formed by the long-range CDW  (see Fig. 3 C). The excess charge within these regions will be screened by the electrons which are now transferred to the delocalised bands.
At a sufficiently low T and high concentration $n_v$, these voids are expected to aggregate by diffusion into domain walls. The overall state becomes conducting via the band states released by the annihilated polarons which, if ordered, would  form a new incommensurate structure.
 We can also imagine that photo-excited electrons could squeeze into the structure in between the polarons creating interstitials with a  concentration $n_i$ ({\it 9}).
Together with voids with a concentration $n_v$ the total "intrinsic defect" concentration $n_d=n_v-n_i$ may have either sign. Overall charge conservation  $n_h+n_v=n_e+n_i$ gives the imbalance of the current carriers $n_d=n_e-n_h$. 
 Conventionally, photodoping is a transient effect, so once $e-h$ symmetry is recovered, the voids and domain walls disappear and the \emph{C} state is restored. However, if the voids can be stabilised by collectively ordering into a long-range ordered state, the final state is different than the original one.

Overall charge conservation requires that the concentration $n_v$ of the voids or their walls compensates for the imbalance  $n_e \ne n_h$  of electron and hole concentrations in delocalized bands, arresting their mutual recombination and maintaining a metallic state.
In either case, local strain causes subsequent self-reorganisation of these voids (or extra electrons), giving rise to a long-range ordered structure with an excess of charge carriers with respect to the \emph{C} state.

Now we outline a plausible minimal model for the metastable state, which may have phenomenological implications beyond our detailed speculations on the nature of microscopic processes in this particular material. The model details, the calculation and more results are presented in the Appendix.

The free energy $F_d(n_d)$  appropriate for the formation of the charge-ordered state outlined above ({\it 16}) needs to include the effect of repulsion between the domain walls,  their crossings ({\it 16,22,23}), and should reproduce the first order nature of the transition ({\it 16,23}). The $F_d(n_d)$ based on these considerations  and existing models ({\it 16,23}) is plotted in Fig. 3 B.

 To obtain the time dependencies of concentrations $n_e(t)$, $n_h(t)$ and of the electronic temperature $T(t)$ we consider the recombination rates of $e$ and $h$ across the spectral gap and into the new ordered state. Apart from the densities, the rates depend on the separate chemical potentials $\mu_{i}(t)=\partial F_i/\partial n_{i}$ for the electrons, holes and defects (we assume that the sub-systems are more or less equilibrated internally ({\it 19})).  The difference in chemical potentials gives the energy released when particles are exchanged among the reservoirs which determines the temperature evolution. In equilibrium, the $\mu_i$ for all the three reservoirs must be equal, so the intersection of the three surfaces $\mu_{e}$, $\mu_{h}$ and $\mu_{d}$, as functions of $n_{e}$ and $n_{h}$ and subject to the constraint $n_d = n_h-n_e$ gives the thermodynamically stable end points of the system. These are shown in Fig. 3 D,  plotted for an arbitrary intermediate temperature $0<T<T_{H}$. The condition $\mu_{h}=\mu_{d}=-\mu_{e}$ gives two stable points, which correspond to the free energy minima shown in Fig. 3 B: point \emph{C} where $n_{e}=n_{h}$, and point \emph{H}, where  $n_e-n_h=n_d\ne 0$. The third intersection \emph{U} is unstable since it corresponds to a maximum of the free energy.

The relaxation processes between the charge reservoirs considered by the model  are shown schematically in Fig. 3 C (See the Appendix for more details).  In Figs. 4 A - C we show the calculated trajectories of $n_{e}(t)$ and $n_{h}(t)$ after a laser pulse excitation plotted as a function of time $t$ in the form of parametric plots for different cycles.
The plots  include the lines of intersections between the pairs of chemical potential surfaces from Fig.3 D,  indicating the possible end points (\emph{H} or \emph{C})  corresponding to the temperature at the end of the cycle.  In the W cycle, we start in the \emph{C} state at low \emph{T} where $n_{e}=n_{h}\simeq 0$ (point O). For excitation above threshold $U_{W}>U_{T}$ (Fig. 4 A),  the laser pulse causes  the electronic temperature to increase and the system trajectory initially follows the \emph{C} state line (where $n_{e}=n_{h}$), then makes a loop and ends  at point \emph{H} (where $n_{d}\ne 0$). In this cycle, the system trajectory goes around the energy maximum point \emph{U} behind which the \emph{H} point is hidden.

Below threshold ($U_{W}<U_{T}$), the system initially follows the same path, but the final temperature reached by the system is  too low for there to be an intersection of all three chemical potentials, so the system returns back to state  \emph{C} (Fig. 4 B).
The appearance of a switching threshold observed in the experiments  is thus described by the model. In the E cycle we start from the hidden state at point \emph{H} (Fig. 4 C). Increasing $T$ above a critical value causes the system to follow curve {\bf 1} towards the commensurate state \emph{C}$^{\prime}$ at an elevated temperature $T>T_{H}$.
Cooling thereafter causes it to return via {\bf 2} to the stable point \emph{C}, reproducing the cycle performed in our experiments.

We can now also understand why switching does not occur for $\tau> 4$ ps: Electron-hole asymmetry may be present  as long as the entire electronic system is out of equilibrium with the lattice (up to $3\sim5$ ps). With pulses longer than a critical length  which is related to the electron/hole energy relaxation time,  the \emph{H} state can no longer form.
Changing the W pulse length $\tau$ within our model and plotting the threshold  $U_{T}$ as a function of pulse length $\tau$, we obtain the curve shown in Fig. 2 C.
The model closely follows the observed behaviour, whereby no stable \emph{H} point is reachable for pulses longer than a critical value $\tau_{c}$, observed experimentally at $\sim 4$ ps.

The main nontrivial observations, namely the appearance of a switching threshold for the W pulse fluence, its critical pulse-length dependence, the threshold temperature for the E cycle,  and the high conductance can thus be reproduced. Moreover, the narrow AM spectrum is a direct consequence of the predicted homogeneous  long range order in the H state. The observation of a mid-gap spectral feature following sub-threshold photoexcitation as reported recently({\it 19,24,25}) is consistent with a transient change of polaron density.

The reason for the remarkable stability of the \emph{H} state over an arbitrarily long time is that it is topologically protected:
The domain wall density $n_d$ cannot change continuously, but can do so only in discrete steps, only when the number of periods of the electronic crystal changes, which is quantized. Such an effect as has been demonstrated in 1D CDWs  ({\it 26,27}). This constraint - and its energetically costly resolution by proliferation of topological plane and line defects  - resembles the protection and inhibited decay of super-currents in superfluids and superconductors where the intrinsic defects are phase slips and vortices.
 Here, on a microscopic level, proliferation of new domain walls provides a mechanism for spectral flow of particles across the spectral gaps
 which converts  polaronic states to/from band states.
 The creation and motion of such extended objects will be substantially slowed down by the presence of intersecting discommensurations and finally arrested by pinning to lattice defects.
The E cycle may be explained by the creep of extended defects which is known to be promoted by heating above an irreversibility line characteristic of pinning phenomena.

Previous time-resolved experiments in \emph{1T}-TaS$_{2}$ failed to detect switching because they were performed either with insufficient fluence({\it 18}), or the ambient temperature was too high({\it 20}).
The mechanism for the creation of a topologically protected state by a nonequilibrium quench is generic and is likely to be found in other charge-ordered materials with multiple electronic reservoirs.
The switching is caused by relatively weak and short pulses, which - considering the large change in resistance and optical reflectivity - has obvious application potential. The effect also opens the way to the search for new generations of room temperature non-volatile memory elements in electronically ordered materials. As a memory element, switchable by sub-35 fs pulses, our device is already comparable to, or exceeds the current speed record of 40 fs in magnetic materials ({\it 28}).

\section{Appendix: the calculation of the system trajectory}
The theory was developed on a basis of a phenomenological approach which leads to a minimal and computationally treatable model, and which is compatible with existing microscopic pictures. Our model is independent of underlying microscopic details and can be only complicated further, but this was not necessary as shown by the successful numerical solutions of the equations below.

The focus is upon three reservoirs of electrons: electrons and holes as mobile charge carriers with concentrations $n_e$ and $n_h$ and the crystallized electrons with the concentration $1-n_d$ where $n_d=n_v-n_i$ is the relative density of intrinsic defects (interstitials "i" and voids "v" as explained above; both are known to be present ({\it 9}). Altogether, the concentrations are subject to the charge conservation law $n_{e}-n_{h}=n_{v}-n_{i}=n_{d}$.
The crystalline density, hence the concentration of its defects, can be changed only by proliferation of topological defects, akin to vortices in superfluids, which is allowed at high T of initial dynamics but is arrested at low T of the already formed H state.

The total free energy, which we shall consider additive,
$F(n_{d},n_{e},n_{h})\approx F_{d}(n_{d})+F(n_{e})+F(n_{h})$
determines the partial chemical potentials
$\mu_{j}=\partial F/\partial n_{j}\approx\mu_{j}(n_{j})$
(these are also functions of the temperature) of the reservoirs which must be all equilibrated in the static regime.
The electrons and the holes from the nominally empty and filled band states will be considered as particles with 2D spectra $\epsilon_{e,h}(p)=\Delta_{e,h}+p^{2}/2m_{e,h}$ characterized by their activation energies - the gaps $\Delta_{e,h}$ and by effective masses $m_{e,h}$ giving constant densities of states $N_{e,h}\sim m_{e,h}$ above the respective gaps. That yields the chemical potentials as

\begin{equation}
\mu_{e,h}(n)=\Delta_{e,h}+k_{B}T\ln(\exp(n_{e,h}/(k_{B}TN_{e,h})-1).
\label{mu-eh}
\end{equation}

The necessary free energy $F_{d}(n_{d})$ of the crystalline reservoir can be taken just to satisfy the fact of the first order phase transition in equilibrium between the $C$ and $IC$ phases, which is the experimental fact as well as the results of thermodynamical calculations. We went a bit further to justify the chosen form of $F_{d}$ by referring to a general theory ({\it 23}) of weakly incommensurate triangular lattices.
Assuming the symmetry between vacancies and interstitials, i.e. with respect to the sign of $n_d$, we choose the parametrization

\begin{equation}
F_{d}(n_{d})=E_{DW}(C_{0}|n_{d}|+C_{1}|n_{d}|e^{-1/(\xi|n_{d}|)}
-C_{2}\xi n_{d}^{2}+C_{4}\xi^{3}n_{d}^{4})\label{eq:F-d}
\end{equation}
where $C_{n}$ are numeric constants. Here $\xi$ is the domain wall width, $E_{DW}$ is its energy scale per a constituent defect. The first two terms are standard for a picture of the 2nd order $C$-$IC$ transition: in thermodynamic equilibrium, the coefficient $C_{0}<1$, as a function of $T$, reduces the DW energy and for $C_{0}<0$ the walls start to be created but their concentration is stabilized by the repulsion given by the second term $\sim C_{1}$.
The next term $\sim C_{2}$ appears for non-collinear arrays of domain walls
which now intersect in points with a concentration $\sim n_{d}^{2}$.
A key point ({\it 23}) is that this energy is expected to be negative, which we took into account with the sign "-" of the term $\sim C_{2}$ in (\ref{eq:F-d}). Together with the last stabilizing term $\sim C_{4}$ to take into account the repulsion between the crossings, we obtain the desired non-monotonous curve for $F_{d}$ shown in Fig. 3A. The chemical potential of the crystalline reservoir is then:

\begin{equation}
\mu_{d}(n_{d})=E_{DW}(C_{0}+C_{1}(1+\frac{1}{\xi|n_{d}|})e^{-1/(\xi|n_{d}|)}
-2C_{2}\xi|n_{d}|+4C_{4}(\xi|n_{d}|)^{3})\mathrm{sign}(n_{d}).
\label{mu-d}\end{equation}

The three surfaces of chemical potentials $-\mu_{e}$, $\mu_{h}$ and $\mu_{d}$ as a function of $n_{e}$, $n_{h}$ and $n_d=n_e-n_h$ respectively are shown in Fig. 3B.
($\mu_h$ and $\mu_d$, as both counting the electron's deficiencies, should be always taken with an opposite sign with respect to $\mu_e$.)

Mutual transformations among the reservoirs, together with the concomitant heat production, are dictated by imbalances of three partial chemical potentials $\mu_j$. The corresponding kinetic equations are chosen in a simplest form satisfying to the condition that the exchange rate among any two reservoirs vanishes when the corresponding chemical potentials become equal $\delta\mu_{i,j}=0$.
To model the nonequilibrium evolution, we need to consider the relaxation kinetics between the three reservoirs.

The rates $R_{ij}$ of particles' exchange among the reservoirs may be complicated functions of $n_j$, $\mu_j$, and $T$ and we need to make physically motivated assumptions. It is common, from physics of semiconductors to gapful correlated systems,  to take the bi-particle form of the e-h recombination $R_{eh}\sim n_e n_h$. The linear relations $R_{hd}\sim n_h$ and $R_{ed}\sim n_e$ imply that the band particles can transform themselves into defects without meeting another particle, e.g. the holes can annihilate with a main part of polarons rather neglecting the small concentration of defects.  In principle the bi-modal parts $R_{ed}^{bm}\sim n_e n_d$ and $R_{hd}^{bm}\sim n_e n_d$ can be also present which we shall not consider here keeping the minimalistic approach. Next we use the most general principle that, as functions of potentials mismatches, $R_{ij}$ changes the sign passing through zero when the potentials coincide: $\delta\mu_{i,j}=0$. Again, we take it into account in a simplest form of the linear dependence $R_{ij}\sim\delta\mu_{i,j}$.
Finally, kinetic equations for the time evolution of $n_{h}(t)$ and $n_{e}(t)$ acquire the form:

\begin{eqnarray}
\frac{dn_{h}}{dt}=-k_{eh}n_{e}n_{h}(\mu_{e}+\mu_{h})-k_{hd}n_{h}(\mu_{h}-\mu_{d}) & +P(t)
\label{eq:rates1}
\\
\frac{dn_{e}}{dt}=-k_{eh}n_{e}n_{h}(\mu_{e}+\mu_{h})-k_{ed}n_{e}(\mu_{e}+\mu_{d}) & +P(t)
\label{eq:rates2}
\end{eqnarray}
where $k_{ij}$ are the coefficients of the recombination rates $R_{i,j}$ after extracting dependencies on $n_{i,j}$ and $\delta\mu_{i,j}$, $P(t)$ is the temporal profile of particles production.
As a justification, notice a distant resemblance of these equations with basic ones describing the photo-voltaic devices (see e.g. ({\it 29,30}); superficially, $n_{d}$ can be compared with a part of electrons trapped by impurities or lattice defects.

The temperature evolution is taken into account by the energy balance equation
\begin{equation}
J=C_{T}dT/dt=\sum_{j}\mu_{j}(dn_{j}/dt),
\label{eq:temp}
\end{equation}
where $C_{T}$ is the heat capacity.
The formulas (\ref{eq:rates1},\ref{eq:rates2}) and (\ref{eq:temp}) form a complete set of equations governing the system evolution. Their solution yields the trajectory of the Fig. 3, and the partial time dependencies in Fig. \ref{fig:Densities}.

In these equations, the time is in ps and $k_{ij}$ are in ps$^{-1}$, the laser pulse length is $\tau=0.035$ ps. Other quantities are dimensionless: $n_j$  as concentrations per enlarged unit cell of one star; the energies as given in units of the electrons' gap $\Delta_{e}$. The parameters used for calculation presented in the main text and in Fig. \ref{fig:Densities} below are as follows.The hole gap is taken as $\Delta_{h}=1.4$ (following a prevailing information from  experiments and band calculations that the Fermi level is closer to the upper rim of the gap), and the densities of states are $N_{h}=1.2$, $N_{e}=1$. For electrons there may be an uncertainty because we do not know for sure if they thermalize to the bottom of the conduction band or to the upper Hubbard level which positions look to be close, see ({\it 31}) and rfs. therein. Also the electronic states are not accessible to photo-emission experiments. Still, it is known from STS ({\it 32}) and optics ({\it 33}) that the total gap is $\Delta_{e}+\Delta_{h}=0.6eV$, so our energy unit is 0.25eV.

Parameters in Eqs. (\ref{eq:F-d},\ref{mu-d}) are adjusted such that the free energy $F_{d}$ shows the appropriate minimum as shown in Fig. 3A: $C_{0}=0.22$, $C_{1}=1$, $C_{2}=1$ and $C_{4}=2$. The domain wall size is $\xi=3$, in lattice units, and the domain wall energy scale is $E_{DW}=2$. So the energy to initiate one defect is taken as
$E_d=C_0 E_{DW}=0.44\Delta_{e}=0.11eV$.
$k_{eh}=20$ ps$^{-1}$, $k_{hd}=20$ ps$^{-1}$ and $k_{ed}=$10 ps$^{-1}$; their magnitudes are estimated from the observed single particle relaxation rate in Fig. 2B. The trajectory is very robust with respect to parameters $k_{ij}$, changing their magnitude by a factor of 2 has a minor effect on the plots, provided that $k_{hd}/k_{ed}>2$.  For these parameters, the threshold laser pulse energy for switching to the $H$ state is $U_{T}=0.8$.

The temporal evolution of $n_{e},$ $n_{h}$, $n_{d}$ and $T$ is plotted in Fig. \ref{fig:Densities} above and below $U_{T}$, corresponding to the trajectories
shown in Fig. 3. Notice that the temperature saturates quite fast while the evolution of concentrations keeps a longer time towards H state and much longer time towards the C state.

\bibliographystyle{Science}

We acknowledge discussions with L. Forro, V. V. Kabanov, N. Kirova,
P. Monceau, E. Tossatti and E. Tutis. Samples were grown by P. Sutar
and H. Berger. We also acknowledge funding from ARRS, European restructuring
funds (CENN Nanocenter) and the ERC advanced grant TRAJECTORY.

\begin{itemize}
\item[1.]
S. Koshihara, et al., 
{\it Phys. Rev. B} \textbf{42}, 6853 (1990).
\item[2.]K. Nasu, {\it Photoinduced phase transitions}, (World Scientific,
2004).
\item[3.]H. Okamoto et al., 
{\it Phys. Rev. B} \textbf{70}, 165202 (2004).
\item[4.]A. Cavalleri, et al., 
{\it Phys. Rev. Lett.} \textbf{87}, 237401 (2001).

\item[5.]S.Tomimoto, S. Miyasaka \& T. Ogasawara, 
{\it Phys Rev B} \textbf{68}, 035106 (2003).

\item[6.] N. Takubo et al. 
{\it Phys. Rev. Lett.} \textbf{95}, 017404 (2005); K. Nasu, H. Ping and H.
Mizouchi, 
{\it J. Phys.: Condens. Matter} \textbf{13}, R693 (2001).

\item[7.] G. Yu, et al., 
{\it Phys. Rev. Lett.} \textbf{67}, 2581 (1991); D. Fausti, et al., 
{\it Science} \textbf{331}, 189 (2011).

\item[8.]A. Zakery and S.R. Elliott, {\it Optical nonlinearities in chalcogenide glasses and their applications.},
New York: Springer (2007).

\item[9.]R. Thomson, B. Burk, A. Zettl, and J, Clarke,
{\it Phys Rev B} \textbf{49}, 16899\textendash{}16916 (1994).

\item[10.]B. Sipos, et al., 
{\it Nat. Mater.} \textbf{7}, 960 (2008).

\item[11.]L.J. Li, et al. 
{\it Eur. Phys. Lett.} \textbf{97}, 67005 (2012 ).

\item[12.]T. Ishiguro and H. Sato, {\it Phys. Rev. B.} 44, 2046
(1991)

\item[13.]J. Demsar, L. Forro, H. Berger, \& D. Mihailovic,
{\it Phys Rev B} \textbf{66}, 041101 (2002).

\item[14.]K. Rossnagel \& N. Smith, {\it Phys. Rev. B} \textbf{73}, 073106 (2006) and references therein.

\item[15.] E.Tosatti and P. Fazekas, {\it J. Phys. Colloques} \textbf{37}, C4 (1976).

\item[16.]K. Nakanishi and H. Shiba, {\it J. Phys. Soc. Jpn.} \textbf{43}, 1839 (1977).

\item[17.]J.C. Petersen et al., 
{\it Phys.Rev.Lett.} \textbf{107,} 177401 (2011)

\item[18.]S. Hellmann et al., 
{\it Phys. Rev.Lett.} \textbf{105}, 187401 (2010).

\item[19.]L. Perfetti, et al. 
{\it New J. Phys.} \textbf{10}, 053019 (2008).

\item[20.]M. Eichberger et al, 
{\it Nature} \textbf{468}, 799 (2010).

\item[21.] F. Clerc et al., J. Phys.: Condens. Matter \textbf{19}, 255002 (2007).

\item[22.] W. McMillan, 
Phys Rev B \textbf{14}, 1496Ð1502 (1976).

\item[23.] For a review and refs. see P. Bak, {\it Rep. Prog. Phys.}
\textbf{45}, 597 (1982). 

\item[24.]N. Dean, et al. 
{\it Phys. Rev. Lett.} \textbf{106}, 016401 (2011).

\item[25.]K. Ishizaka et al., 
{\it Phys.Rev.B} \textbf{83}, 081104(R) (2011)

\item[26.] D.V. Borodin, et al, JETP \textbf{66}, 793 (1987) and refs. therein;

\item[27.] S.G. Zybtsev, et al, {\it Nat. Comms.} \textbf{1}, 85 (2010).

\item[28.]A. Kirilyuk, A. Kimel, \& T. Rasing, 
{\it Rev. Mod. Phys.} \textbf{82}, 2731 (2010).

\item[29]J. Piprek, Semiconductor Optoelectronic devices (Academic Press, SanDiego, 2003).

\item[30]S. Selberherr, Analysis and Simulation of Semiconductor Devices, Springer-Verlag, Vienna, Austria, 1984.

\item[31] J.K. Freericks1, H.R. Krishnamurthy, Yizhi Ge, A.Y. Liu \& Th. Pruschke, Phys. Status Solidi B \textbf{246}, 948–954 (2009).

\item[32] Ju-Jin Kim, W. Yamaguchi, T. Hasegawa \& K. Kitazawa,
Phys. Rev. Lett. \textbf{73}, 2103 (1994).

\item[33] L. V. Gasparov and K. G. Brown, A.C. Wint, D.B. Tanner, H. Berger, G. Margaritondo, R. Gaal, and L. Forro, Phys. Rev. B \textbf{66}, 094301 (2002).

\end{itemize}

\clearpage
\begin{figure}[tbh]
\begin{centering}
\includegraphics[width=0.9\textwidth,angle=0]{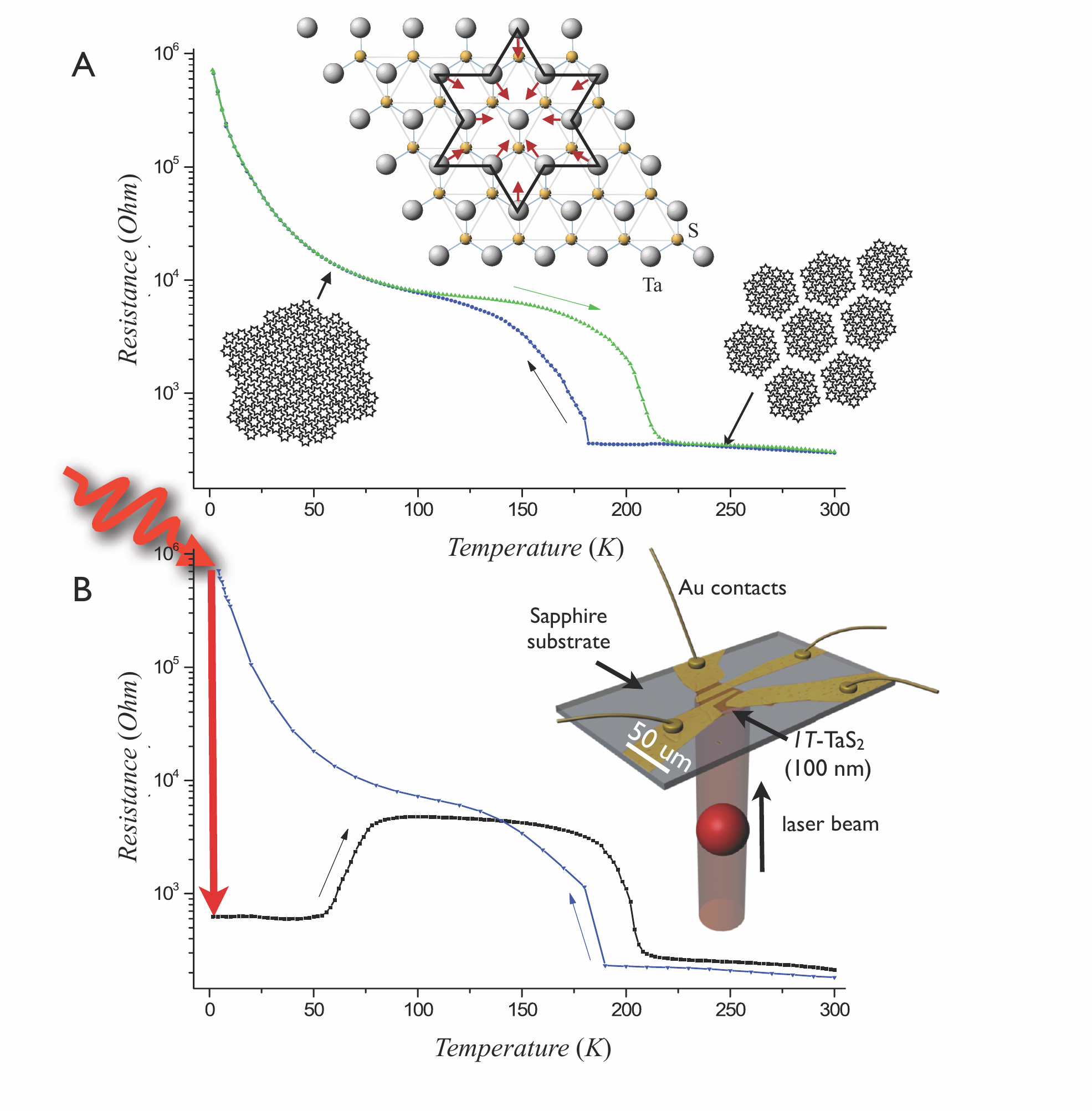}
\par\end{centering}

\caption{
\noindent {\bf Fig. 1.} Resistivity switching of \emph{1T}-TaS$_{2}$ by a 35 fs laser pulse at 800 nm.
(A) The $T$-dependence of the 4-probe resistance \emph{r(T)} on temperature cycling. The inserts show schematically the lattice distortions associated with an individual polaron and their ordering in the \emph{NC} and \emph{C} states respectively.
(B) The drop of \emph{r} at 1.5 K after a single pulse with $U_{w}>U_{T}$ (red arrow). On heating the resistance reverts back between 60 and 100K (black
curve). The insert shows a schematic of the sample and contacts, which
are from an optical microscope image.
}

\end{figure}

\begin{figure}[tbh]
\begin{centering}
\includegraphics[width=0.9\textwidth,angle=0]{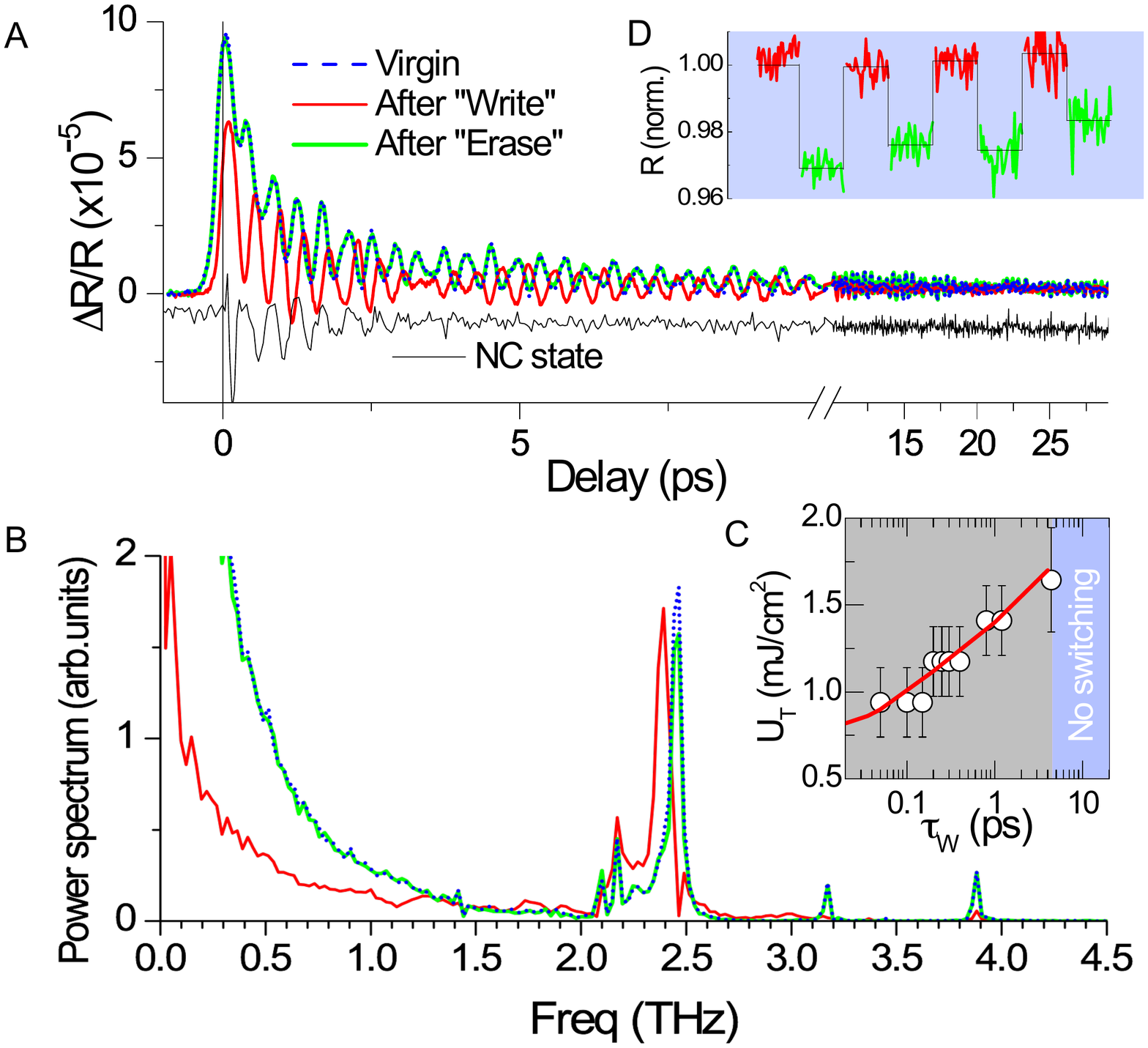}
\par\end{centering}

\caption{
\noindent {\bf Fig. 2.}
(A) The transient reflectivity $\Delta R(t)/R$ of $1T-$TaS$_{2}$
in the virgin state (blue dashed line), after exposure to a 50 fs
W pulse (red line), and after an E pulse (green line). For comparison,
the thin black line is the data in the \emph{NC} state at 220 K recorded
on cooling (the trace is offset for clarity).
(B) The corresponding FT spectra $S(\omega)$ using the same color notation. Note the complete disappearance of the AM at 2.46 THz in the \emph{H} state, and clean switching back after the E pulse. The switching is also observed for
the modes at 2.3, 3.2 and 3.85 THz. The \emph{NC} state spectrum, as well as the \emph{T}-state spectrum recorded at 240 K on heating (not shown) are qualitatively different than either the \emph{C} or \emph{H} state spectra.
(C) The switching threshold fluence $U_{T}$ as a function of pulse length $\tau_{W}$ measured optically with the Pump-probe experiments. The red line is predicted by the model calculation.
(D) The reflectivity at 800 nm recorded with the photodiode during a sequence of alternating W and \emph{E} pulses. (The noise is from the laser.)}
\end{figure}

\begin{figure}[tbh]
\begin{centering}
\includegraphics[width=0.9\textwidth,angle=0]{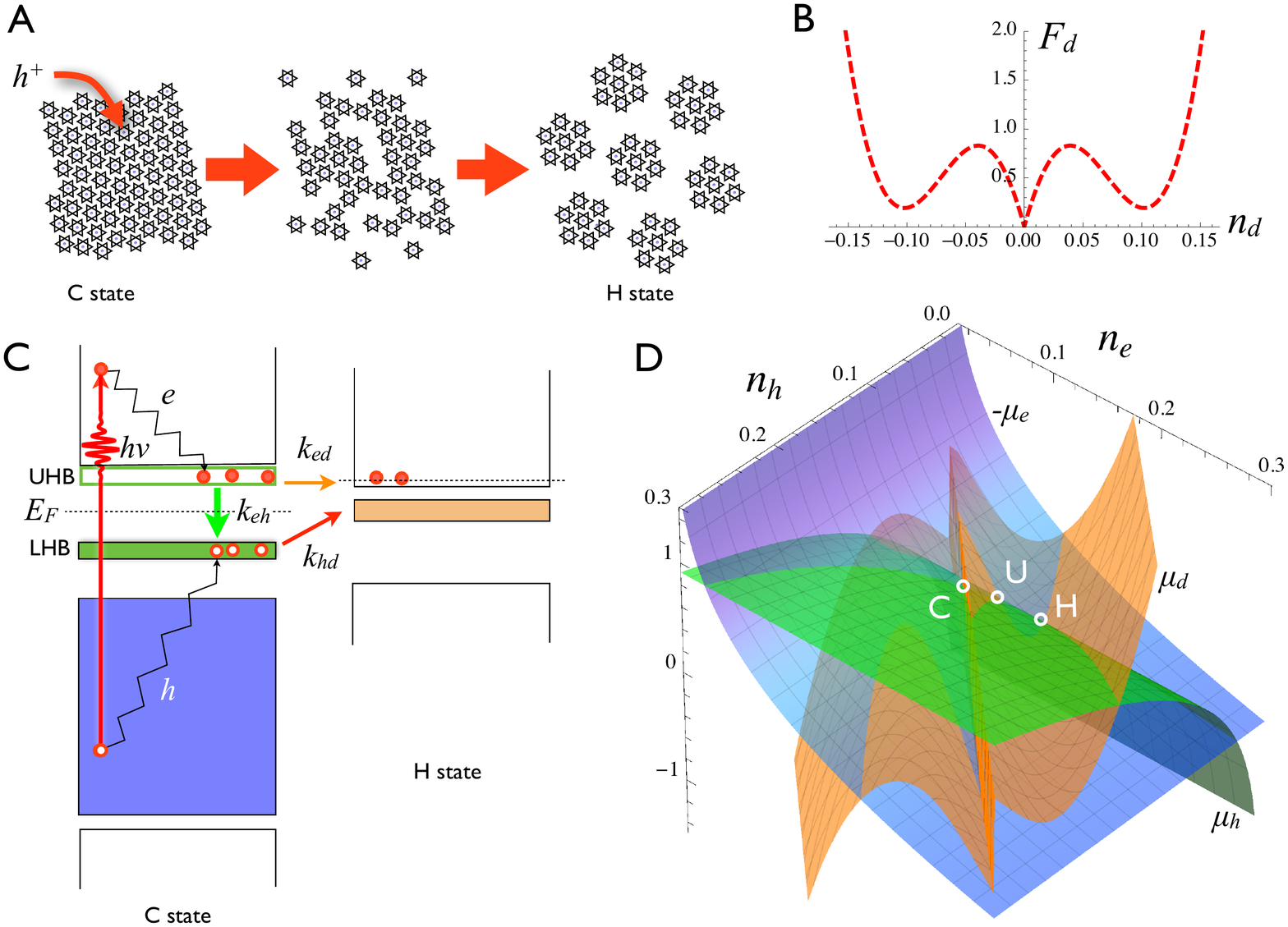}
\par\end{centering}
\caption{
\noindent {\bf Fig. 3.}
(A) A  schematic diagram of the reordering following a laser pulse. First a hole annihilates with a polaron, causing voids in the structure, which reorganise into a long-range ordered structure with a wavevector shift ${\delta q}$ is related to the number of defects by charge conservation ${\delta q}/\pi \simeq n_{d}$. Loosely, the structure may be thought of as polaron clusters separated by domain walls.
(B) The CDW free energy $F_{d}$ (dashed red line) as a function of $n_{d}$.
(C) A schematic energy levels diagram of the C state based on refs. ({\it 14,15,28}). Occupied bands in blue are those from Ta atoms within each polaron in Fig. 1. The Hubbard bands are shown in green. Photoexcitation, initial energy relaxation and subsequent relaxation processes of the $e$ and $h$ amongst themselves and to a new incommensurate H state are shown schematically.
(D) The chemical potential surfaces $-\mu_{e}$ (blue), $\mu_{h}$ (green) and $\mu_{d}$ (orange) respectively as a function of $n_{e}$ and $n_{h}$. The intersections \emph{C} and \emph{H} are commensurate and incommensurate stable points of the system respectively.}
\end{figure}

\begin{figure}[tbh]
\begin{centering}
\includegraphics[width=0.9\textwidth,angle=0]{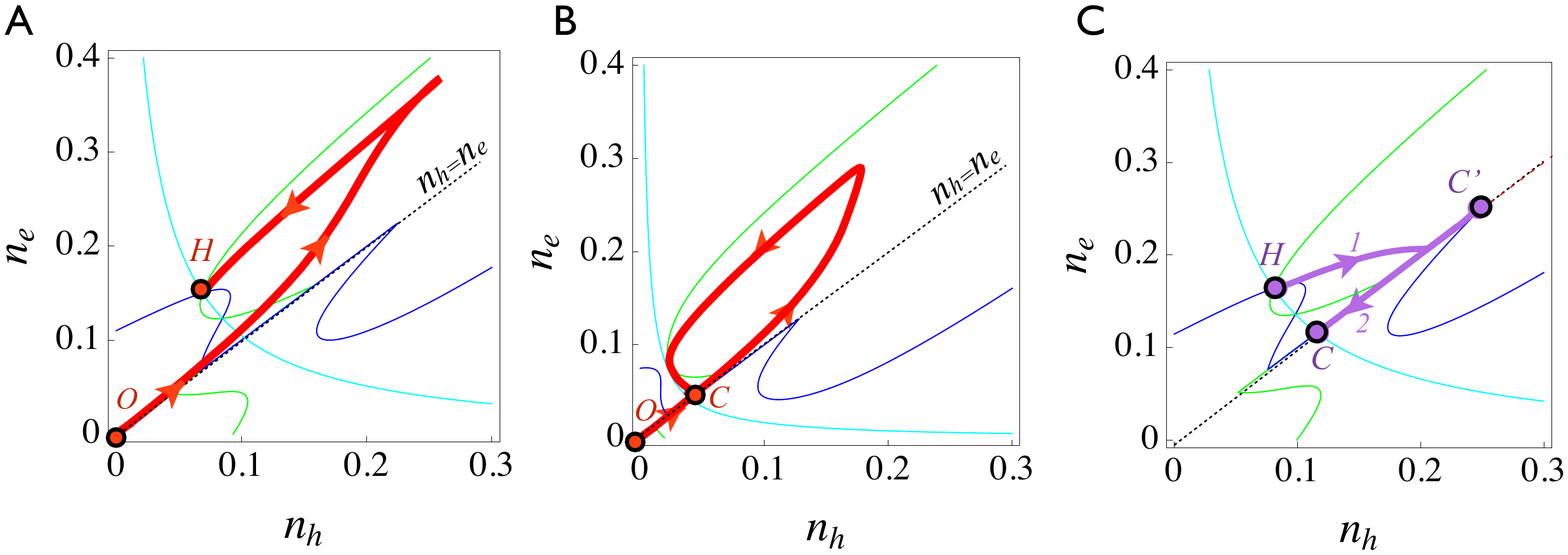}
\par\end{centering}
\caption{
\noindent {\bf Fig. 4.}
(A) The system trajectory system (thick red line) above switching threshold $U_{W}>U_{T}$ on a parametric plot of $n_{e}(t)$ and $n_{h}(t)$) as a function of time superimposed on lines of partial equilibria given by the intersections $\mu_{e}=-\mu_{d}$ (green), $\mu_{h}=\mu_{d}$ (blue) and $\mu_{e}=-\mu_{h}$ (cyan) drawn for the final temperature after the system has stabilized.
(B) Same as (A) except below threshold $U_{W}<U_{T}$. The system now returns to the \emph{C} state.
(C) Model trajectory of erasure by heating in the \emph{H}-state (purple line): the trajectory first leaves the \emph{H}-point, then joins the commensurate line. After reaching a maximum at \emph{C'}, on cooling the system returns
to point \emph{C}.}
\end{figure}

\begin{figure}[tbh]
\includegraphics[width=1\columnwidth]{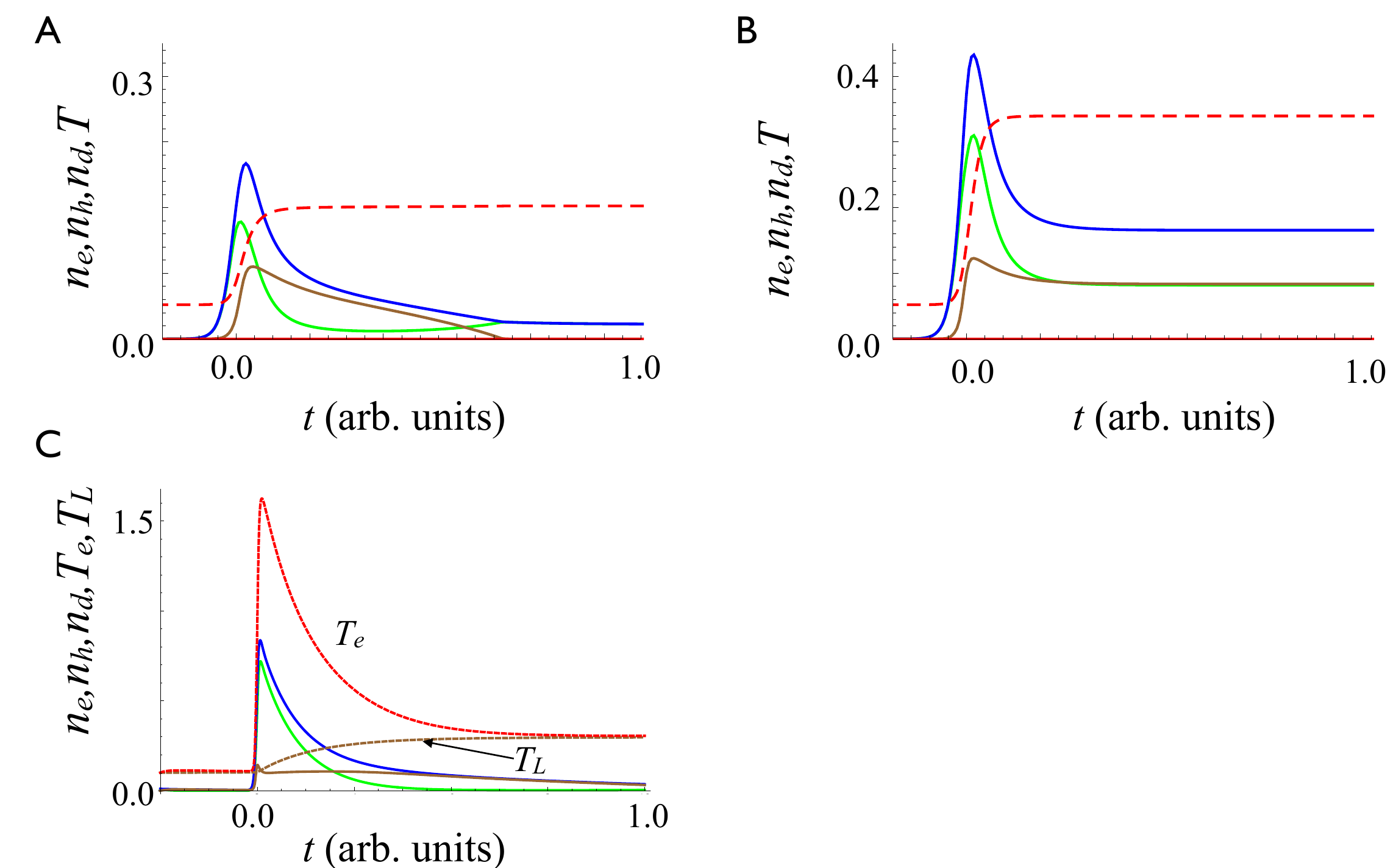}
\caption {The calculated $n_{e}$ (blue), $n_{h}$ (green), $n_{d}$ (brown)
and the temperature $T$ (red dashed) as functions of time using the equations (\ref{eq:rates1},\ref{eq:rates2}) and (\ref{eq:temp}):
a) below threshold ($U_W<U_T$) and b) above threshold ($U_W>U_T$). The initial temperature was taken as $T=0.1 \Delta_{e}/k_{B}$. The corresponding parametric plots of $n_e$ and $n_h$ are shown in Figs. 3 D and E respectively.
Note that the calculation does not discuss the cooling part of the cycle, so the temperature remains high after all the pulse energy is transferred. In C we show the results of a calculation which includes the two-temperature model for $U_W > U_T$. $n_e$ (blue), $n_h$ (green), $n_d$ (brown) are qualitatively similar to the results of the simple model calculation (B). The electronic and lattice temperatures $T_e$ and $T_L$ are shown by the red-dashed and brown-dashed lines respectively.}
\label{fig:Densities}
\end{figure}

\end{document}